\documentclass[11pt]{article}

\usepackage{amsmath}
\usepackage{graphicx}
\usepackage{indentfirst}
\usepackage{amssymb}
\usepackage{cite}
\usepackage{color}
\usepackage{subfigure}

\begin{document}

\title{\bf{Weak Cosmic Censorship Conjecture\\in Kerr-(Anti-)de Sitter Black Hole with Scalar Field}}

\date{}
\maketitle

\begin{center}
\author{Bogeun Gwak}$^a$\footnote{rasenis@sejong.ac.kr}\\

\vskip 0.25in
$^{a}$\it{Department of Physics and Astronomy, Sejong University, Seoul 05006, Republic of Korea}\\
\end{center}
\vskip 0.6in

{\abstract
{We investigate the weak cosmic censorship conjecture in Kerr-(anti-)de Sitter black holes under the scattering of a scalar field. We test the conjecture in terms of whether the black hole can exceed the extremal condition with respect to its change caused by the energy and angular momentum fluxes of the scalar field. Without imposing the laws of thermodynamics, we prove that the conjecture is valid in all the initial states of the black hole (non-extremal, near-extremal, and extremal black holes). The validity in the case of the near-extremal black hole is different from the results of similar tests conducted by adding a particle because the fluxes represent the energy and angular momentum transferred to the black hole during the time interval not included in the tests involving the particle. Using the time interval, we show that the angular velocity of the black hole with the scalar field of a constant state takes a long time for saturation to the frequency of the scalar field.}}

\thispagestyle{empty}
\newpage
\setcounter{page}{1}

\section{Introduction}\label{sec1}

Black holes, which are directly proven to exist through detection by the Laser Interferometer Gravitational-Wave Observatory (LIGO), are among the most interesting topics in gravity theories. Classically, in the black hole spacetime, there is an event horizon through which no matter can escape from the black hole; thus, no radiation from the black hole can reach an observer located outside this horizon. However, in quantum theory, black holes act as thermal systems that emit energy through Hawking radiation\cite{Hawking:1974sw,Hawking:1976de}. In Hawking radiation, the Hawking temperature is defined for a black hole. Furthermore, when a particle is added to the black hole, depending on the conserved quantities of the particle, the conserved quantities of the black hole, such as mass and angular momentum, can increase or decrease. However, an irreducible quantity exists during this process, which is known as the irreducible mass\cite{Christodoulou:1970wf,Bardeen:1970zz,Christodoulou:1972kt}. The irreducible mass is the energy distributed on the surface of the horizon of the black hole \cite{Smarr:1972kt}. Owing to the similarity between the irreducible mass and the thermodynamic entropy, the Bekenstein--Hawking entropy is defined to be proportional to the surface area of the black hole, which is the square of its irreducible mass\cite{Bekenstein:1973ur,Bekenstein:1974ax}. Accordingly, we can establish the laws of thermodynamics for black holes.

The thermodynamic properties of a black hole, such as the Hawking temperature, Bekenstein--Hawking entropy, and thermodynamic potentials, are all defined on its horizon. Thus, without the horizon, the laws of thermodynamics for black holes cannot be defined. In other words, the thermodynamics of a black hole is strongly dependent on the stability of its horizon. Interestingly, the stability of the horizon has been suggested by the weak cosmic censorship conjecture\cite{Penrose:1964wq,Penrose:1969pc}. This conjecture was originally proposed for a stable horizon to prevent the breakdown of the causality at a naked singularity owing to the singularity of the black hole located inside the horizon. Thus, the stability of the horizon is a necessary condition for the validity of the weak cosmic censorship conjecture. To prove the validity of the conjecture, we have to investigate each type of black hole because no general method can be used for such a proof. Thus far, various black holes have been tested by various methods. The first investigation of the conjecture showed its validity for the extremal Kerr black hole by adding a particle\cite{Wald:1974ge}. The cosmic censorship conjecture also depends on the state of black holes. For the Kerr black hole, adding a particle makes the horizon unstable in the near-extremal case\cite{Jacobson:2009kt}. This instability of the horizon can be resolved in consideration of the self-force effect\cite{Barausse:2010ka,Barausse:2011vx,Colleoni:2015ena,Colleoni:2015afa,Sorce:2017dst}. When black holes are coupled with the Maxwell field, the horizon becomes unstable because of overcharging beyond the extremal condition, as in the case of a Reissner--Nordstr\"{o}m black hole\cite{Hubeny:1998ga} in consideration of a backreaction. However, the stability of the horizon depends on the analysis method; hence, a counterexample exists for the overcharging of the Reissner--Nordstr\"{o}m black hole with a backreaction\cite{Isoyama:2011ea}. Thus far, the weak cosmic censorship conjecture has been studied in various black holes by adding a particle\cite{BouhmadiLopez:2010vc,Gwak:2011rp,Rocha:2011wp,Crisostomo:2003xz,Gao:2012ca,Hod:2013vj,Zhang:2013tba,Rocha:2014jma,McInnes:2015vga,Cardoso:2015xtj,Siahaan:2015ljs,Natario:2016bay,Horowitz:2016ezu,Duztas:2016xfg,Revelar:2017sem,Duztas:2017lxk,Song:2017mdx,Liang:2018wzd,Yu:2018eqq}. In such cases, without consideration of a self-force effect or backreaction, the extremal black holes cannot be overcharged, but the near-extremal black holes can be overcharged by the energy carried by the particle. Furthermore, the changes in the black hole are consistent with the laws of thermodynamics; hence, the validity of the conjecture is closely related to the laws of thermodynamics\cite{Gwak:2015fsa,Gwak:2017kkt}. The conjecture can also be investigated by the scattering of test fields, such as scalar and Maxwell fields\cite{Hod:2008zza,Semiz:2005gs,Toth:2011ab,Duztas:2013wua,Semiz:2015pna,Natario:2016bay,Duztas:2017lxk}. For the test fields, the validity of the conjecture depends on the method of investigation, but most cases present results similar to those obtained by adding a particle.

The scattering of an external field to a black hole provides unexpected features compared to those observed by the addition of a particle. One such feature is superradiance, whereby an external field extracts conserved quantities from the horizon of the black hole when it is scattered out from the black hole. For superradiance, the state of the external field defined by $\omega$ and $m$ (i.e., the frequency and azimuthal number, respectively) satisfies the following inequality with respect to the angular velocity of the black hole $\Omega_\text{h}$: $\omega < m \Omega_{h}$\cite{Zeldovich:1971aa,Zeldovich:1972ab}. The energy radiated by superradiance can be amplified or dissipated according to the boundary condition; thus, superradiance is associated with the instability of a system, including black holes in its time evolution\cite{Brito:2015oca} (see references therein). Various studies have investigated the instabilities caused by superradiance. For anti-de Sitter (AdS) cases, large Kerr-AdS black holes are stable under superradiance because of the high resonance frequencies, but the superradiance can be amplified in small Kerr-AdS black holes, making them unstable\cite{Hawking:1999dp,Cardoso:2004hs,Cardoso:2006wa,Uchikata:2009zz}. The propagation of a scalar field has recently been studied at a proper temperature for de Sitter (dS) black holes, such as Schwarzschild-dS and Kerr-dS black holes\cite{Crispino:2013pya,Kanti:2014dxa,Pappas:2016ovo}. In dS black holes, the cosmological horizon is located outside the black holes. The cosmological horizon also emits energy similar to the horizon of the black hole; thus, an observer can detect two temperatures for radiations from the horizon of the black hole and the cosmological horizon. These radiations are not in equilibrium, which is directly related to the stability of the dS black hole, and a proper temperature should be set\cite{Pappas:2017kam}. Although the propagation of a scalar field in (A)dS black holes is complex, various analytical studies have investigated (A)dS black holes\cite{Teukolsky:1973ha,Suzuki:1998vy,Suzuki:1999nn,Cho:2009wf,Yoshida:2010zzb,Dias:2013sdc,Cardoso:2013pza,Zhang:2014kna,Brito:2015oca,Delice:2015zga,Ganchev:2016zag}. In particular, the scalar field non-minimally coupled with gravity has recently been studied in dS black holes\cite{Crispino:2013pya,Kanti:2014dxa,Pappas:2016ovo,Ahmed:2016lou,Pappas:2017kam} because such a scalar field acts differently in a low-frequency mode compared with the case of minimal coupling. This is because non-minimal coupling can be an effective mass in the propagation\cite{Crispino:2013pya}.

In this paper, we investigate the weak cosmic censorship conjecture in Kerr-(anti-)de Sitter (K(A)dS) black holes under the scattering of a scalar field. Starting from the Lagrangian of a scalar field with the non-minimally coupling term in the black hole background with a static boundary, we determine the energy and angular momentum fluxes at the outer horizon. The mass and angular momentum of the black hole are then assumed to change as much as those carried by the fluxes during an infinitesimal time interval. Under the change in the black hole, {\it without imposing the laws of thermodynamics}, we prove that the Bekenstein--Hawking entropy proportional to the horizon area is irreducible in the non-extremal black hole; hence, the conjecture is valid in the non-extremal case. However, this proof is not applicable to the extremal and near-extremal cases for the following reasons: (1) the extremal black hole needs to be analyzed by a different method because of the divergence of the change in the entropy and (2) various near-extremal black holes can be naked singularities by adding a particle without a self-force effect or backreaction\cite{Hubeny:1998ga,Jacobson:2009kt,Colleoni:2015ena}. Considering these two issues, we investigate the infinitesimal change under the fluxes of the scalar field in the component of the metric $g^{rr}$ that governs the locations of the horizons. Remarkably, we can show that the conjecture is valid in the near-extremal (including extremal) black hole. This result differs considerably from the results of the tests on the conjecture by adding a particle without a self-force effect or backreaction. The change caused by the fluxes of the scalar field includes information about the infinitesimal time that is not included in the particle case; hence, it acts as a constraint that prevents the extremal condition from being exceeded. Furthermore, to understand the emission of the scalar field, in the constant state of the scalar field, we show that the angular velocity of the black hole cannot be saturated to the state of the scalar field in a finite time. In other words, the black hole needs an infinitely long time to complete its emission because of the scattering of the scalar field.

The remainder of this paper is organized as follows: Section\,\ref{sec2} introduces the K(A)dS black hole transformed to the metric with a static asymptotic boundary; Section\,\ref{sec3} solves the scalar field equation at the outer horizon of the K(A)dS black hole in a standard manner; Section\,\ref{sec4} proves the laws of thermodynamics in the non-extremal K(A)dS black hole; Section\,\ref{sec5} investigates the weak cosmic censorship conjecture in near-extremal and extremal K(A)dS black holes without imposing the laws of thermodynamics; Section\,\ref{sec6} studies the change in the angular velocity of the black hole in the constant state of the scalar field; and finally, Section\,\ref{sec7} summarizes our results.

\section{Kerr-(Anti-)de Sitter Black Holes}\label{sec2}

The K(A)dS black hole is a solution to Einstein's gravity with a cosmological constant $\Lambda$. We consider herein an arbitrary cosmological constant such that the asymptotic geometries are flat under $\Lambda=0$, anti-de Sitter (AdS) under $\Lambda<0$, and de Sitter (dS) under $\Lambda>0$. The metric of the K(A)dS black hole is well known in the Boyer-Lindquist coordinates as
\begin{align}\label{eq:metric01}
ds^2&=-\frac{\Delta_r}{\rho^2}\left(dt-\frac{a\sin^2\theta}{\Xi} d\phi\right)^2+\frac{\rho^2}{\Delta_r}dr^2+\frac{\rho^2}{\Delta_\theta}d\theta^2+\frac{\Delta_\theta\sin^2\theta}{\rho^2}\left(a\,dt-\frac{r^2+a^2}{\Xi}d\phi\right)^2\,,\\\rho^2&=r^2+a^2\cos^2\theta\,,\,\,\Delta_r=(r^2+a^2)(1-\frac{1}{3}\Lambda r^2)-2Mr\,,\,\,\Delta_\theta=1+\frac{1}{3}a^2\Lambda\cos^2\theta\,,\,\,\Xi=1+\frac{1}{3}a^2\Lambda\,,\nonumber
\end{align}
where the mass and spin parameter are given by $M$ and $a$, respectively. However, in the metric of Eq.\,(\ref{eq:metric01}), the angular velocity is not zero at the asymptotic boundary $r\gg 1$ in the AdS case. This nonzero angular velocity implies that the asymptotic observer is not static, but rotating with respect to the boundary. Furthermore, the first law of thermodynamics is invalid because of the non-static observer in the metric of Eq.\,(\ref{eq:metric01}). The K(A)dS black hole for the static observer is obtained from the coordinate transformation introduced in \cite{Hawking:1998kw}. Then,
\begin{align}\label{eq:coordinatetransformation}
t\rightarrow T, \quad \phi\rightarrow \Phi+\frac{1}{3}a\Lambda T,
\end{align}
by which the asymptotic angular velocity becomes zero. The metric of the K(A)dS black hole is transformed to 
\begin{align}\label{eq:metric02}
ds^2=-\frac{\Delta_r}{\rho^2\Xi^2}\left(\Delta_\theta dT-a\sin^2\theta d\Phi\right)^2+\frac{\rho^2}{\Delta_r}dr^2+\frac{\rho^2}{\Delta_\theta}d\theta^2+\frac{\Delta_\theta\sin^2\theta}{\rho^2\Xi^2}\left(a\left(1-\frac{1}{3}\Lambda r^2\right)dT-(r^2+a^2)d\Phi\right)^2,
\end{align}
where the angular velocity at the outer horizon $r_\text{h}$ is
\begin{eqnarray}\label{eq:angularvelocity07}
\Omega_\text{h}=\frac{a\left(1-\frac{1}{3}\Lambda r_\text{h}^2\right)}{r_\text{h}^2+a^2}\,.
\end{eqnarray}
The first law of thermodynamics is well defined under the choice of the angular velocity\cite{Caldarelli:1999xj,Gibbons:2004ai}. Note that the second law of thermodynamics and the weak cosmic censorship conjecture also become valid in the metric in Eq.\,(\ref{eq:metric02}) under particle absorption\cite{Gwak:2015fsa}. The mass and the angular momentum of the K(A)dS black hole, $M_\text{B}$ and $J_\text{B}$, respectively, are given by\cite{Hawking:1998kw,Caldarelli:1999xj,Gibbons:2004ai}
\begin{eqnarray}\label{massdefined01}
M_\text{B}=\frac{M}{\Xi^2}\,,\quad J_\text{B}=\frac{Ma}{\Xi^2}\,,
\end{eqnarray}
where we set $G=1$. In this work, we mainly consider the behaviors of the K(A)dS black hole under the scattering of a scalar field at the outer horizon and the weak cosmic censorship conjecture. These behaviors are closely related to the laws of thermodynamics at the outer horizon at which the scalar field is scattered. The Hawking temperature and the Bekenstein-Hawking entropy at the outer horizon are
\begin{align}\label{eq:ent07}
T_\text{h}=\frac{r_\text{h}\left(1-\frac{a^2}{r_\text{h}^2}-\frac{a^2\Lambda}{3}-r_\text{h}^2\Lambda\right)}{4\pi\left(r_\text{h}^2+a^2\right)},\quad  S_\text{h}=\frac{1}{4}\mathcal{A}_\text{h}=\frac{\pi\left(r_\text{h}^2+a^2\right)}{\Xi},
\end{align}
where the surface of the outer horizon is $\mathcal{A}_\text{h}$. The first law of thermodynamics is given by
\begin{align}\label{eq:1stlaw01}
dM_\text{B}=T_\text{h}dS_\text{h}+\Omega_\text{h} dJ_\text{B}.
\end{align}
In addition, for the KdS black hole, another horizon, called the cosmological horizon (denoted by $r_\text{c}$), exists because of the positive cosmological constant. The temperature and the entropy at the cosmological horizon can be defined in the same way as the temperature and entropy at the outer horizon. The first law of thermodynamics at the cosmological horizon is expressed as\cite{Dolan:2013ft,Kubiznak:2015bya,Kubiznak:2016qmn}
\begin{align}\label{eq:1stlawcosmologicalh}
dM_\text{B}=-T_\text{c}dS_\text{c}+\Omega_\text{c} dJ_\text{B},\quad T_\text{c}=\frac{r_\text{c}\left(1-\frac{a^2}{r_\text{c}^2}-\frac{a^2\Lambda}{3}-r_\text{c}^2\Lambda\right)}{4\pi\left(r_\text{c}^2+a^2\right)},\quad  S_\text{c}=\frac{\pi\left(r_\text{c}^2+a^2\right)}{\Xi}.
\end{align} 
Note that we focus on the outer horizon of the K(A)dS black hole with a general value of the cosmological constant. Thus, we do not use Eq.\,(\ref{eq:1stlawcosmologicalh}).

\section{Solution to Scalar Field Equation}\label{sec3}

We investigate the infinitesimal changes in the K(A)dS black hole when the external energy and the angular momentum are transferred from the scattering of a complex scalar field. The transferred energy and the angular momentum are obtained from the energy and angular momentum fluxes of the scalar field at the outer horizon. We need to determine the energy-momentum tensor $T_{\mu\nu}$ of the solution to the scalar field at the outer horizon. The Lagrangian for the complex scalar field in the K(A)dS background is
\begin{align}\label{eq:scalarlag01}
S_\Psi =-\frac{1}{2}\int d^4 x \sqrt{-g}\left(\partial_\mu \Psi \partial^\mu \Psi^*+(\mu^2+\xi R)\Psi\Psi^*\right),
\end{align}
where $R$ denotes the curvature; $\mu$ denotes the mass of the scalar field; and $\xi$ denotes the non-minimal coupling constant. We consider herein the Lagrangian of the massive scalar field with a non-minimal coupling constant as a general case of a scalar field. However, as shown below, the effects from the mass and non-minimal coupling terms are removed from the solution of the scalar field at the outer horizon. The equations of motion then become
\begin{align}\label{eq:eom10}
\frac{1}{\sqrt{-g}} \partial_\mu \left(\sqrt{-g} g^{\mu\nu} \partial _\nu \Psi\right)-(\mu^2+\xi R) \Psi=0,\quad  \sqrt{-g}=\frac{\rho^2 \sin\theta}{\Xi}.
\end{align}
In the K(A)dS black hole, the scalar field equation given by Eq.\,(\ref{eq:eom10}) is separable\cite{Vasudevan:2004ca,Page:2006ka}. We take the solution of the scalar field as
\begin{align}\label{eq:solution03}
\Psi(T,r,\theta,\Phi)=e^{-i\omega T}e^{im\Phi} R(r) \Theta(\theta).
\end{align}
Imposed on the solution in Eq.\,(\ref{eq:solution03}), the equation of motion is separated under a constant $\lambda$ into
\begin{align}
\label{eq:radialeq01}&\frac{1}{R}\frac{d}{dr}\left(\Delta_r \frac{d}{dr} R\right)-\frac{1}{\Delta_r}\left(-\omega^2 (r^2+a^2)^2 + 4Mar\omega m -\frac{1}{9}a^2 (-3+r^2\Lambda)^2 m^2\right)-(\mu^2+\xi R)r^2 =\lambda,\\
\label{eq:thetaeq01}&-\frac{1}{\sin\theta}\frac{1}{\Theta}\frac{d}{d\theta}\left(\sin\theta\Delta_\theta \frac{d}{d\theta}\right)\Theta + \left(\frac{a^2 \sin^2\theta}{\Delta_\theta} \omega^2 +\Delta_\theta \csc^2\theta m^2 +(\mu^2+\xi R)a^2 \cos^2\theta\right)=\lambda,
\end{align}
where the non-minimal coupling term $\xi R$ plays the same role as the mass because the background spacetime is fixed. Thus, the equations of motion are similar to those of the massive scalar field with minimal coupling in the K(A)dS black hole. We mainly focus herein on the solution of the radial equation given by Eq.\,(\ref{eq:radialeq01}). When energy and angular momentum fluxes are obtained at the limit of the outer horizon, the radial solution plays an important role; otherwise, the contribution of the $\theta$-directional solution in Eq.\,(\ref{eq:thetaeq01}) will be reduced to unity in the fluxes by the normalization condition: $\int \Theta^2(\theta)d\Omega=1$. In addition, the solution of the $\theta$-directional equation has been treated and discussed in various studies\cite{Teukolsky:1973ha,Suzuki:1998vy,Suzuki:1999nn,Cho:2009wf,Yoshida:2010zzb,Dias:2013sdc,Cardoso:2013pza,Zhang:2014kna,Brito:2015oca,Delice:2015zga,Ganchev:2016zag,Pappas:2017kam} (see also references therein). In several limits, we can expect the form of $\Theta(\theta)$. In the non-rotating case, $a=0$, the $\theta$-directional equation becomes that of the Schwarzschild-dS black hole; hence, $\lambda=\ell (\ell+1)$ of the angular momentum number\cite{Pappas:2017kam}. In the case of a Kerr black hole, $\Lambda=0$, the equation corresponds to the well-known spheroidal equation\cite{Teukolsky:1973ha}. In the general case of a K(A)dS black hole, these radial and $\theta$-directional equations reduce to Heun's equations, and should be solved by a numerical method\cite{Suzuki:1998vy,Suzuki:1999nn,Cho:2009wf,Yoshida:2010zzb}. The eigenvalue of $\lambda$ is still close to the angular momentum, but is not an integer\cite{Delice:2015zga}. In the slowly rotating case, $a\omega \ll 1$, the exact form of $\lambda$ is given as a series expansion\cite{Suzuki:1999nn,Cho:2009wf,Zhang:2014kna}. As we have already shown herein, the detailed form of $\Theta(\theta)$ does not contribute to the fluxes at the outer horizon. Thus, we omit the determination of a solution to the $\theta$-directional equation.

To solve the radial equation given by Eq.\,(\ref{eq:radialeq01}), we define a tortoise coordinate as
\begin{align}
\frac{dr^*}{dr}=\frac{(r^2+a^2)}{\Delta_r},
\end{align}
where the intervals of $r$ and $r^*$ coordinates are distinct with respect to the signs of $\Lambda$. For $\Lambda=0$, the interval $(r_\text{h},+\infty)$ of the $r$ coordinate becomes $(-\infty,+\infty)$ of the $r^*$ coordinate. For $\Lambda<0$, the interval $(r_\text{h},+\infty)$ of the $r$ coordinate becomes $(-\infty,0)$ of the $r^*$ coordinate. For $\Lambda>0$, the interval $(r_\text{h},r_\text{c})$ of the $r$ coordinate becomes $(-\infty,+\infty)$ of the $r^*$ coordinate. These provide various boundary conditions at $r\rightarrow +\infty$ depending on $\Lambda$. We focus on the outer horizon, and we can take a common boundary condition at $r\rightarrow r_\text{h}$. The radial equation is rewritten in terms of the $r^*$ coordinate as
\begin{align}
&\frac{d^2 R}{d {r^*}^2}+\frac{2r\Delta_r}{(r^2+a^2)^2}\frac{d R}{d {r^*}}+\left(\omega^2 - \frac{4Mar\omega m}{(r^2+a^2)^2}+\frac{m^2 a^2}{(r^2+a^2)^2}\left(1-\frac{1}{3}r^2\Lambda\right)^2-\frac{\Delta_r}{(r^2+a^2)^2}\left((\mu^2+\xi R) r^2 +\lambda\right)\right)\nonumber\\&=0,\nonumber
\end{align}
which, at the limits of the outer horizon, $r\rightarrow r_\text{h}$ and $\Delta_r\rightarrow 0$, becomes
\begin{align}
\frac{d^2 R}{d {r^*}^2}+\left(\omega -\frac{a m}{2Mr_\text{h}}\left(1-\frac{1}{3}r^2\Lambda\right)\right)^2=0.
\end{align}
Therefore, the solution of the radial equation at the outer horizon is obtained with Eq.\,(\ref{eq:angularvelocity07}) as
\begin{align}\label{radialsol21}
R(r)\sim e^{\pm (\omega - m \Omega_\text{h})r^*},
\end{align} 
where we take the solution with a negative sign as the ingoing wave to the outer horizon. The scalar field solution and its conjugate are
\begin{align}\label{eq:solution07}
\Psi=e^{-i\omega T} e^{im \Phi} e^{-i (\omega - m \omega_\text{h})r^*} \Theta(\theta),\quad \Psi^*=e^{i\omega T} e^{-im \Phi} e^{i (\omega - m \omega_\text{h})r^*} \Theta^*(\theta).
\end{align}

\section{Thermodynamics under Scattering of Scalar Field}\label{sec4}

In the scattering, the conserved quantities of the K(A)dS black hole change because of the energy and angular momentum carried by the scalar field. When the scalar field carries a sufficiently large angular momentum to overspin the black hole, the horizon of the black hole will disappear. The weak cosmic censorship can be violated in this case. We investigate herein whether the changes in the K(A)dS black hole are consistent with the laws of thermodynamics in the scattering of the scalar field. The changes in the black hole are estimated in the infinitesimal time interval $dT$; hence, the changes in the energy and angular momentum are also infinitesimal because of the time interval. Furthermore, during the time interval, the changes in the energy and angular momentum are too small to consider other interactions, such as a self-force effect. Therefore, our analysis provides the first-order change in the black hole, which is the most important in the time interval.

The energy and angular momentum carried by the scalar field can be estimated from their fluxes at the outer horizon. The fluxes can be obtained from the energy-momentum tensor derived from the Lagrangian of the scalar field in Eq.\,(\ref{eq:scalarlag01}). The energy-momentum tensor is\begin{align}
T^\mu_\nu&=\sum_i\frac{\partial \mathcal{L}}{\partial (\partial_\mu \Psi^i)}\partial_\nu \Psi^i -\delta^\mu_\nu\mathcal{L}\\
&=\frac{1}{2}\partial^\mu\Psi \partial_\nu \Psi^*+\frac{1}{2}\partial^\mu\Psi^* \partial_\nu \Psi-\delta^\mu_\nu\left(\frac{1}{2}\partial_\mu\Psi \partial^\mu\Psi^* -\frac{1}{2}(\mu^2+\xi R)\Psi\Psi^*\right).\nonumber
\end{align}
Carried by the scalar field, the energy and angular momentum fluxes are given as
\begin{align}\label{eq:flux01}
\frac{dE}{dT}=\int T_T^r \sqrt{-g}d\theta d\Phi,\quad \frac{dL}{dT}=-\int T^r_\Phi \sqrt{-g}d\theta d\Phi.
\end{align}
When we take the solutions of Eq.\,(\ref{eq:solution07}), the fluxes in Eq.\,(\ref{eq:flux01}) indicate that the energy and angular momentum flow into the outer horizon of the black hole. Coming into the outer horizon, the energy and angular momentum of the scalar field cannot be distinct from those of the black hole; hence, they can be assumed to be absorbed into the black hole. The corresponding conserved quantities of the black hole (i.e., mass and angular momentum) change as much as the fluxes of the scalar field. During the time interval $dT$, the changes in the mass and angular momentum of the black hole become
\begin{align}\label{eq:fluxes07}
dM_\text{B}=\left(\frac{dE}{dT}\right)dT,\quad dJ_\text{B}=\left(\frac{dL}{dT}\right)dT,
\end{align}
where
\begin{align}\label{eq:fluxes05}
\frac{dE}{dT}=\omega(\omega-m\Omega_\text{h}) \frac{(r_\text{h}^2+a^2)}{\Xi},\quad \frac{dL}{dT}=m(\omega-m\Omega_\text{h}) \frac{(r_\text{h}^2+a^2)}{\Xi}=\frac{m}{\omega}\frac{dE}{dT}.
\end{align}
Note that we use the normalization condition of $\Theta(\theta)$ in the integration with respect to the solid angle. The fluxes in Eq.\,(\ref{eq:fluxes05}) represent the amount of energy and angular momentum flowing into the outer horizon. However, in the case of $\omega < m\Omega_\text{h}$, the signs of the fluxes become negative. The negative sign implies that the energy and the angular momentum flow out from the black hole; thus, the scattering of the scalar field acts as energy extraction from the black hole. This is called superradiance\cite{Zeldovich:1971aa,Zeldovich:1972ab}, which is a process of a field similar to the Penrose process of a particle\cite{Bardeen:1970zz,Penrose:1971uk}. The actual condition for superradiance is that the boundary condition of the scalar field needs to be in the asymptotic region. In contrast to the outer horizon, the transmission rate of the scalar field can be nonzero at the asymptotic boundary. For example, for the KdS black hole, superradiance occurs at $m\Omega_\text{c}<\omega < m\Omega_\text{h}$ because of the cosmological horizon $r_\text{c}$\cite{Khanal:1986cu,Zhang:2014kna}. However, we focus herein on the infinitesimal change in the K(A)dS black hole caused by the energy and angular momentum fluxes at the outer horizon during the infinitesimal time interval $dT$. Thus, the analysis results do not depend on the asymptotic boundary conditions.

In the scattering of the scalar field, the change in the configuration of the black hole can be represented in terms of $\omega$ and $m$ of the field characteristics without imposing the laws of thermodynamics or the weak cosmic censorship conjecture. When the final state is assumed for a K(A)dS black hole, the initial black hole of $(M_\text{B},J_\text{B},r_\text{h})$ becomes the final black hole of $(M_\text{B}+dM_\text{B},J_\text{B}+dJ_\text{B},r_\text{h}+dr_\text{h})$ from Eq.\,(\ref{eq:fluxes07}). The change in the location of the outer horizon is then obtained from $g^{rr}\sim\Delta_r (M_\text{B},J_\text{B},r_\text{h})=0$
\begin{align}
\Delta_r(M_\text{B}+dM_\text{B},J_\text{B}+dJ_\text{B},r_\text{h}+dr_\text{h})=\frac{\partial \Delta_r}{\partial M_\text{B}}\Big{|}_{r=r_\text{h}}dM_\text{B}+\frac{\partial \Delta_r}{\partial J_\text{B}}\Big{|}_{r=r_\text{h}}dJ_\text{B}+\frac{\partial \Delta_r}{\partial r}\Big{|}_{r=r_\text{h}}dr_\text{h}=0,
\end{align}
where
\begin{align}
\frac{\partial \Delta_r}{\partial M_\text{B}}\Big{|}_{r=r_\text{h}}&=\frac{8}{3}a^2 r_\text{h} \Lambda \Xi - 2r_\text{h}\Xi^2-\frac{2a^2 \left(1-\frac{r_\text{h}^2 \Lambda}{3}\right)\Xi^2}{M},\quad
\frac{\partial \Delta_r}{\partial J_\text{B}}\Big{|}_{r=r_\text{h}}=-\frac{8}{3}a r_\text{h} \Lambda \Xi +\frac{2a \left(1-\frac{r_\text{h}^2 \Lambda}{3}\right)\Xi^2}{M},\nonumber\\
\frac{\partial \Delta_r}{\partial r}\Big{|}_{r=r_\text{h}}&=-2M-\frac{2}{3}r_\text{h}(r_\text{h}^2+a^2)\Lambda+2 r_\text{h} \left(1-\frac{1}{3}r_\text{h}^2\Lambda\right)\equiv d\Delta_\text{h}.\nonumber
\end{align}
The initial location of the outer horizon $r_\text{h}$ satisfies $\Delta_r(M_\text{B},J_\text{B},r_\text{h})=0$. After time $dT$, the mass and the angular momentum of the black hole change because of the fluxes of the scalar field; hence, based on the changes in the mass and angular momentum, the final location of the outer horizon will satisfy $\Delta_r(M_\text{B}+dM_\text{B},J_\text{B}+dJ_\text{B},r_\text{h}+dr_\text{h})=0$. The change in the outer horizon $dr_\text{h}$ then becomes
\begin{align}\label{eq:radious01}
dr_\text{h}=\frac{(r_\text{h}^2+a^2)(3 a m\Xi-a m r_\text{h}\Lambda (4M+r_\text{h}\Xi)-3Mr_\text{h}\Xi \omega + a^2\omega(-3\Xi +r_\text{h}\Lambda (4M+r_\text{h}\Xi)))(\omega-m\Omega_\text{h})}{M(3M+r_\text{h}(-3+a^2\Lambda + 2r_\text{h}^2\Lambda))}dT,
\end{align}
where we impose Eqs.\,(\ref{eq:fluxes07}) and (\ref{eq:fluxes05}). The change in the location of the outer horizon $dr_\text{h}$ can be negative or positive depending on our choice of parameters $\omega$ and $m$ of the scalar field. The Bekenstein-Hawking entropy changes as follows:
\begin{align}
dS_\text{h}=\frac{\partial S_\text{h}}{\partial M_\text{B}}dM_\text{B}+\frac{\partial S_\text{h}}{\partial J_\text{B}}dJ_\text{B}+\frac{\partial S_\text{h}}{\partial r_\text{h}}dr_\text{h},
\end{align}
in which
\begin{align}
\frac{\partial S_\text{h}}{\partial M_\text{B}}=\frac{2a^2\pi (r_\text{h}^2+a^2)\Lambda}{3M}-\frac{2a^2 \pi \Xi}{M},\quad
\frac{\partial S_\text{h}}{\partial J_\text{B}}=-\frac{2a\pi (r_\text{h}^2+a^2)\Lambda}{3M}+\frac{2a \pi \Xi}{M},\quad
\frac{\partial S_\text{h}}{\partial r_\text{h}}=\frac{2\pi r_\text{h}}{\Xi}.\nonumber
\end{align}
The change in the entropy then becomes
\begin{align}\label{eq:entropy02}
d S_\text{h}=\frac{d S_\text{h}}{dT}dT=\frac{4\pi (r_\text{h}^2+a^2)^2}{d\Delta_\text{h} \Xi}(\omega - m \Omega_\text{h})^2dT\geq 0,
\end{align}
where $d\Delta_\text{h}\equiv \frac{\partial \Delta_r}{\partial r}\big{|}_{r=r_\text{h}}$. At the location of the outer horizon, $d\Delta_\text{h}\geq 0$, and the equality is at the extremal black hole; thus, the entropy or surface area of the non-extremal black hole is irreducible in the scattering of the scalar field. Therefore, the changes in the non-extremal black hole follow well the second law of thermodynamics in general values of the cosmological constant $\Lambda$. That is, even if the energy and the angular momentum are carried or extracted by a scalar field, the surface of the outer horizon covering the singularity still exists for the non-extremal black hole. Furthermore, by the infinitesimal contribution in the interval $dT$, the energy or angular momentum is too small to overspin the non-extremal black hole; hence, the non-extremal black hole might be impossible in this infinitesimal contribution of the scalar field. However, this proof is not complete. We still need to investigate the case of the near-extremal and extremal black holes, where Eq.\,(\ref{eq:entropy02}) is divergent. In addition, the extremal black hole is in the state saturating the extremal condition; hence, overspinning the black hole beyond extremality by an extremely small transferred energy or angular momentum is possible. This will be investigated in the next section.

The irreducible entropy given by Eq.\,(\ref{eq:entropy02}) implies that the scattering of the scalar field follows the second law of thermodynamics. We show that the first law of thermodynamics can be derived from Eq.\,(\ref{eq:entropy02}). Eq.\,(\ref{eq:entropy02}) can be rewritten as
\begin{align}
d S_\text{h}=\frac{4\pi (r_\text{h}^2+a^2)}{d\Delta_\text{h}}(dM_\text{B}-\Omega_\text{h}dJ_\text{B}),
\end{align}
which, by inserting the Hawking temperature $T_\text{h}$ in Eq.\,(\ref{eq:ent07}), becomes
\begin{align}\label{eq:1stlaw03}
dM_\text{B}=T_\text{h}d S_\text{h}+\Omega_\text{h}dJ_\text{B}.
\end{align}
This is the first law of thermodynamics in general values of the cosmological constant $\Lambda$, as expected in Eq. \,(\ref{eq:1stlaw01}). Note that our proof about laws of thermodynamics is applicable to general values of $\Lambda$ and consistent with \cite{Gao:2001ut} in $\Lambda=0$ and \cite{Natario:2016bay} in $\Lambda\leq 0$. We emphasize herein that the first law of thermodynamics is related to the change between the black hole states; hence, changes, such as $dM_\text{B}$, $dJ_\text{B}$, and $dS_\text{h}$, are all defined between the black hole states in Eq.\,(\ref{eq:1stlaw03}). We assume a non-extremal black hole and an infinitesimal change; therefore, the laws of thermodynamics can be obtained in Eqs.\,(\ref{eq:entropy02}) and (\ref{eq:1stlaw03}). In particular, before Eq.\,(\ref{eq:1stlaw03}), Eq.\,(\ref{eq:entropy02}) already ensures the weak cosmic censorship conjecture; hence, obtaining Eq.\,(\ref{eq:1stlaw03}) bears no contradiction. However, we must not impose the laws of thermodynamics to investigate the weak cosmic censorship conjecture in the near-extremal, including extremal, K(A)dS black hole. The near-extremal black hole is a saturated solution close to the extremal condition; hence, the angular momentum is already maximum for a given mass of the black hole. The near-extremal black hole can be overspun by adding an infinitesimal angular momentum, and it becomes a naked singularity with no horizon. Therefore, the first law of thermodynamics is not defined between a black hole and a naked singularity because the final state is not a black hole in this case. Furthermore, the second law of thermodynamics is singular because $d\Delta_\text{h}=0$ in Eq.\,(\ref{eq:entropy02}) for the extremal black hole. Without imposing the laws of thermodynamics, we investigate the weak cosmic censorship conjecture by the scattering of the scalar field in the near-extremal K(A)dS black hole.

\section{Weak Cosmic Censorship Conjecture in Near-Extremal and\\Extremal K(A)dS Black Holes}\label{sec5}

The near-extremal K(A)dS black hole, including extremal case, is in a state in which its angular momentum is nearly saturated to its mass; hence, overspinning the black hole is possible by adding a small portion of the angular momentum, which makes the black hole a naked singularity exceeding the extremal condition. No horizon exists in the naked singularity; hence, the laws of thermodynamics are well defined in this case. Therefore, we cannot impose the laws of thermodynamics. To estimate the final state, we focus on the behaviors of the function $\Delta_r$, which remains well defined in the $g^{rr}$ component, regardless of whether the metric in Eq.\,(\ref{eq:metric02}) represents a black hole or a naked singularity under the scattering of the scalar field. We determine whether a range of $\omega/m$ of the scalar field exists to overspin the near-extremal black hole. When the scattering of the scalar field overspins the near-extremal black hole, the function $\Delta_r$ has no solutions corresponding to the inner and outer horizons in the naked singularity. This can be proven by the sign of the minimum value of the function $\Delta_r$ under the change caused by the scattering. The minimum point is located between the inner and outer horizons; thus, the minimum value of $\Delta_r$ is slightly negative in the near-extremal K(A)dS black hole of the initial state. The change in $\Delta_r$ will change the minimum value because of the scattering. The minimum value can be freely assumed to be infinitesimally small enough to compete with the change in the black hole because of the scattering; hence, the final state can be a naked singularity. We can estimate the final state and the existence of the horizons from the sign of the minimum value. For the positive minimum value, no solutions correspond to the inner and outer horizons; hence, the final state becomes a naked singularity. Otherwise, for the negative minimum value, inner and outer horizons exist; therefore, the final state is still a black hole.

\begin{figure}[h]
\centering
\subfigure[{$\Delta_r$ graph in $\Lambda=0$.}] {\includegraphics[scale=0.6,keepaspectratio]{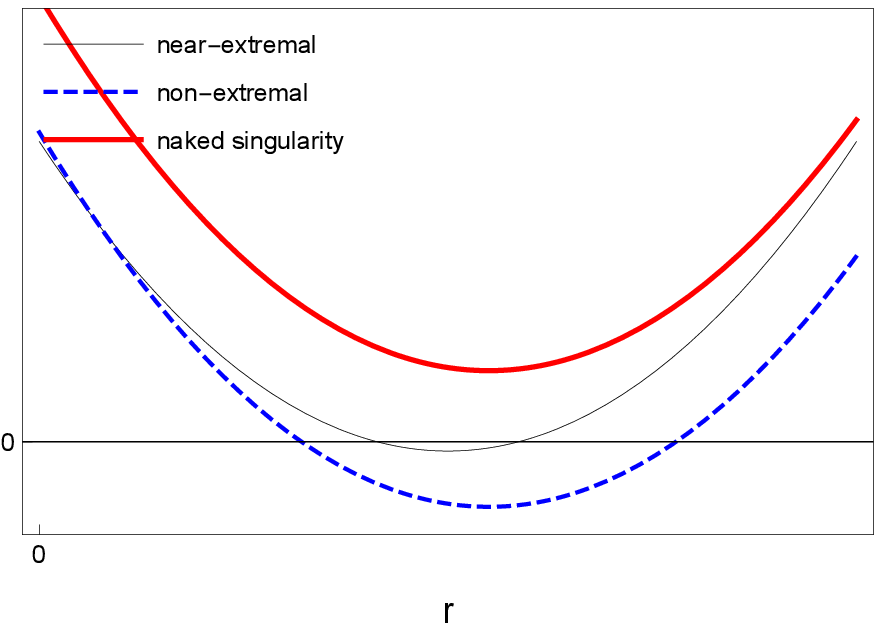}}\quad
\subfigure[{$\Delta_r$ graph in $\Lambda<0$.}] {\includegraphics[scale=0.6,keepaspectratio]{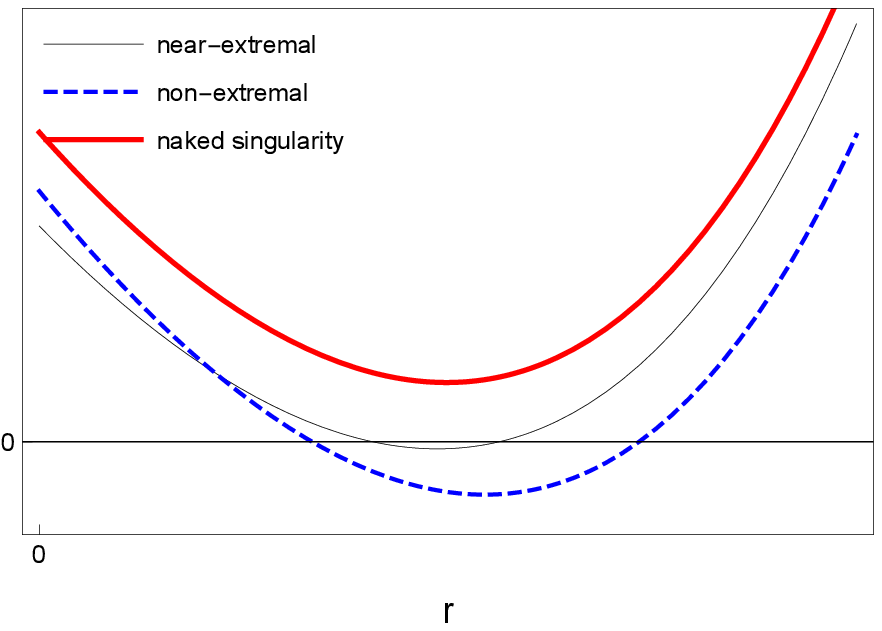}}\quad
\subfigure[{$\Delta_r$ graph in $\Lambda>0$.}] {\includegraphics[scale=0.6,keepaspectratio]{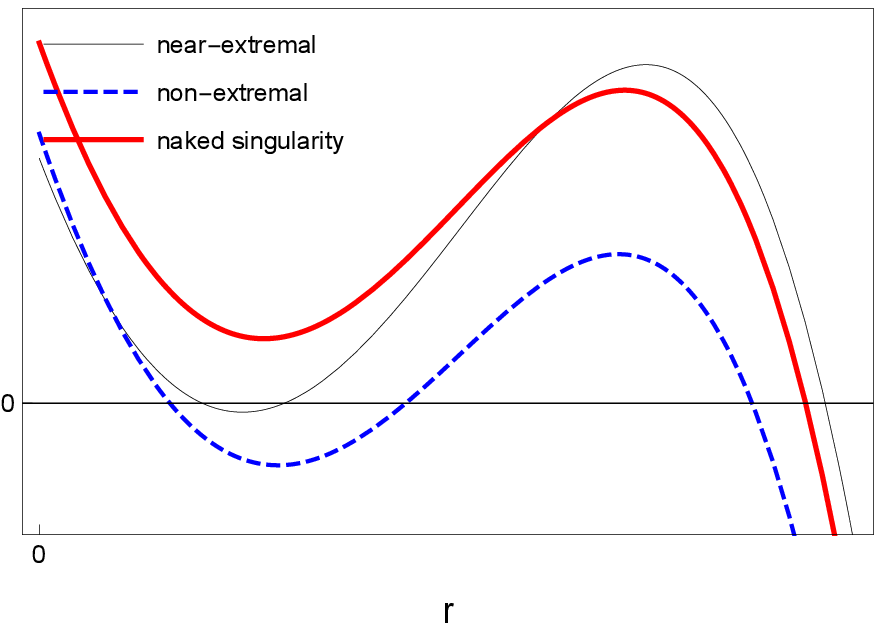}}
\caption{{\small $\Delta_r$ graphs for near-extremal, non-extremal, naked singularity cases for given cosmological constants.}}
\label{fig:fig1}
\end{figure}

The asymptotic boundary depends on the sign of the cosmological constant; thus, the detailed behaviors differ, but the overall conditions are similar to each other, as shown in Fig.\,\ref{fig:fig1}.
In the initial state, as shown by the black lines, the horizons of the initial near-extremal black hole for $(M_\text{B},J_\text{B})$ are located beside the minimum point. The scattering of the scalar field changes the mass and the angular momentum to $(M_\text{B}+dM_\text{B},J_\text{B}+dJ_\text{B})$. Consequently, the minimum location and the value of the function $\Delta_r$ change as shown by the red and blue lines, respectively. For the naked singularity, the function $\Delta_r$ has a positive minimum value of the red lines because of overspinning. The negative minimum values of the blue lines still have horizons of the black hole. Therefore, we investigate the sign of the minimum value under changes in the scattering. Note that another solution in Fig.\,\ref{fig:fig1} is the cosmological horizon $r_\text{c}$ originating from the positive cosmological constant, which is not involved in our investigation of the weak cosmic censorship conjecture.

In terms of the function $\Delta_r$, the minimum condition with the near-extremal black hole in the initial state of $(M_\text{B},J_\text{B})$ is given by
\begin{align}\label{eq:cosmicinitial01}
\Delta_r(M_\text{B},J_\text{B},r_\text{min})\equiv \Delta_\text{min}&=(r_\text{min}^2+a^2)(1-\frac{1}{3}\Lambda r_\text{min}^2)-2Mr_\text{min}\equiv\delta,\\
\frac{d\Delta_\text{min}}{dr_\text{min}}&=-2M-\frac{2}{3}r_\text{min}(r_\text{min}^2+a^2)\Lambda+2 r_\text{min} \left(1-\frac{1}{3}r_\text{min}^2\Lambda\right)=0,\nonumber\\
\frac{d^2\Delta_\text{min}}{dr_\text{min}^2}&=-\frac{8r_\text{min}^2\Lambda}{3}-\frac{2}{3}(r_\text{min}^2+a^2)\Lambda+2\left(1-\frac{1}{3}r_\text{min}^2\Lambda\right)>0,\nonumber
\end{align}
where we define $\Delta_\text{min}$ for simplicity without confusion, and the minimum value $\delta$ is negative and infinitesimally small, $\delta<0$, and $|\delta|\ll 1$. In addition, at the location of the outer horizon $r_\text{h}$, $\Delta_r(M_\text{B},J_\text{B},r_\text{h})$=0. The minimum point infinitesimally moves to $r_\text{min}+dr_\text{min}$ because of the scattering of the scalar field, and $\Delta_r$ is a function of $(M_\text{B}+dM_\text{B},J_\text{B}+dJ_\text{B})$. For these infinitesimal changes, the minimum value of the function $\Delta_r$ becomes
\begin{align}
\Delta_r(M_\text{B}+dM_\text{B},J_\text{B}+dJ_\text{B},r_\text{min}+dr_\text{min})=\Delta_\text{min}+\frac{\partial \Delta_\text{min}}{\partial M_\text{B}}dM_\text{B}+\frac{\partial \Delta_\text{min}}{\partial J_\text{B}}dJ_\text{B}+\frac{\partial \Delta_\text{min}}{\partial r_\text{min}}dr_\text{min},
\end{align}
where
\begin{align}
\frac{\partial \Delta_\text{min}}{\partial M_\text{B}}&\equiv d\Delta_\text{M}=\frac{8}{3}a^2 r_\text{min} \Lambda \Xi - 2r_\text{min}\Xi^2-\frac{2a^2 \left(1-\frac{1}{3}r_\text{min}^2 \Lambda\right)\Xi^2}{M},\nonumber\\
\frac{\partial \Delta_\text{min}}{\partial J_\text{B}}&\equiv d\Delta_\text{J}=-\frac{8}{3}a r_\text{min} \Lambda \Xi +\frac{2a \left(1-\frac{1}{3}r_\text{min}^2 \Lambda\right)\Xi^2}{M}.\nonumber
\end{align}
The changed minimum value can be different from its initial value $\delta$. By imposing the energy and angular momentum fluxes and initial condition in Eqs.\,(\ref{eq:flux01}) and (\ref{eq:cosmicinitial01}), the change in the minimum value is obtained as
\begin{align}\label{eq:deltamin02}
\Delta_r(M_\text{B}+dM_\text{B},J_\text{B}+dJ_\text{B},r_\text{min}+dr_\text{min})&=\delta+d\Delta_\text{M}dM_\text{B}+d\Delta_\text{J}dJ_\text{B}\\
&=\delta+m^2\frac{(r_\text{h}^2+a^2)}{\Xi}d\Delta_\text{M}\left(\frac{\omega}{m}-\Omega_\text{h}\right)\left(\frac{\omega}{m}-\Omega_\text{eff}\right)dT,\nonumber
\end{align}
where the effective angular velocity $\Omega_\text{eff}$ is crucial to the sign of the minimum value in the final state. With Eq.\,(\ref{eq:cosmicinitial01}),
\begin{align}
\Omega_\text{eff}&\equiv\left(-\frac{\partial \Delta}{\partial J_\text{B}}\Big{/}\frac{\partial \Delta}{\partial M_\text{B}}\right)
=\frac{a(4 r_\text{min}^2\Lambda (2 r_\text{min}^2 \Lambda +a^2 \Lambda -3)-3(r_\text{min}^2\Lambda -3)\Xi)}{4a^2 r_\text{min}^2 \Lambda (2 r_\text{min}^2 \Lambda +a^2 \Lambda -3)-3(r_\text{min}^2+a^2)(2r_\text{min}^2 \Lambda -3)\Xi}\\
&=\frac{(-3+r_\text{min}^2\Lambda)(-3+a^2\Lambda+2r_\text{min}^2\Lambda)a+6a\Lambda \delta}{-3(r_\text{min}^2+a^2)(-3+a^2\Lambda+2r_\text{min}^2\Lambda)+6a^2\Lambda \delta}.\nonumber
\end{align}
As $\Omega_\text{eff}$ is written in terms of $r_\text{min}$, we need to rewrite $\Omega_\text{eff}$ with $r_\text{h}$ for comparison with $\Omega_\text{h}$. Under the near-extremal condition in Eq.\,(\ref{eq:cosmicinitial01}), the outer horizon is extremely close to the minimum point, as shown in Fig.\,\ref{fig:fig1}. When we assume the infinitesimal distance $\epsilon$ between the minimum point and the outer horizon, the initial minimum value $\delta$ can be rewritten in terms of $\epsilon$. Then,
\begin{align}\label{eq:epsilondefine03}
r_\text{h}\equiv r_\text{min}+\epsilon,\quad\delta=\left(-1+\frac{a^2\Lambda}{3}+2r_\text{h}^2\Lambda\right)\epsilon^2+\mathcal{O}(\epsilon^3),
\end{align}
where $\epsilon\geq 0$. Thus, the effective angular velocity is obtained in terms of $r_\text{h}$ and $\epsilon$. Finally,
\begin{align}\label{eq:solomegaeff03}
\Omega_\text{eff}=\Omega_\text{h}+\frac{2a r_\text{h} \Xi}{(r_\text{h}^2+a^2)^2}\epsilon + \mathcal{O}(\epsilon^2),
\end{align}
where the signs of $\Omega_\text{eff}$ and $\Omega_\text{h}$ are coincident. For the general value of the cosmological constant $\Lambda$, the effective angular velocity is slightly larger than $\Omega_\text{h}$ with respect to $a>0$; hence, $\omega/m$ of the scalar field can obtain the positive minimum value. The solution to $\Delta_\text{min}+d\Delta_\text{min}>0$ is for the naked singularity
\begin{align}\label{eq:inequaleq01}
\left(\frac{\omega}{m}\right)^2-(\Omega_\text{eff}+\Omega_\text{h})\left(\frac{\omega}{m}\right)+\Omega_\text{eff}\Omega_\text{h}+\frac{\Xi\delta}{m^2(r_\text{h}^2+a^2)d\Delta_\text{M}dT}<0,
\end{align}
where we use $d\Delta_\text{M}<0$. Here, two parameters are assumed to be on the infinitesimal scale: $\epsilon$ and $dT$. Actually, we freely define $\epsilon$ such that we can set $\epsilon\sim dT$. The inequality in Eq.\,(\ref{eq:inequaleq01}) has no solution because the discriminant becomes
\begin{align}
-\frac{4\Xi\left(-1+\frac{a^2\Lambda}{3}+2r_\text{h}^2\right)}{m^2 d\Delta_\text{M}}\epsilon+\frac{4}{9}r_\text{h}\left(\frac{a^2r_\text{h}\Xi^2}{9(r_\text{h}^2+a^2)^4}+\frac{24\Lambda \Xi}{m^2 d\Delta_\text{M}}\right)\epsilon^2+\mathcal{O}(\epsilon^3)<0,
\end{align}
which is negative in the first order of $\epsilon$. Thus, under the scattering of the scalar field, the near-extremal K(A)dS black hole cannot be overspun. In addition, the final state depends on $d\Delta_\text{min}$ in Eq.\,(\ref{eq:deltamin02}). The final black hole becomes more non-extremal than the initial one for any state of $\omega$ and $m$ in the scalar field. In other words, the energy transferred from the scalar field is greater than the angular momentum transferred from it. For the equality $(\omega/m)=\Omega_\text{h}$ or $(\omega/m)<\Omega_\text{eff}$, the change in the minimum value $d\Delta_\text{min}$ also becomes zero; hence, the initial and final states are identical.

Note that one of our results is consistent with that of a previous study on {\it extremal} black holes involving the addition of a test particle or test fields\cite{Toth:2011ab,Gao:2012ca,Natario:2016bay,Gwak:2016gwj,Sorce:2017dst}. When $\epsilon=0$, the initial state becomes the extremal K(A)dS black hole in Eq.\,(\ref{eq:epsilondefine03}) and $\Delta_\text{min}=0$. {\it Without imposing the laws of thermodynamics}, our result implies that the extremal K(A)dS black hole with $\epsilon=0$ cannot be overspun owing to the scattering by solving the scalar field equation. This result is conclusively consistent with that of a previous study on the Kerr black hole with $\Lambda=0$\cite{Sorce:2017dst} and {\it extremal} KAdS black hole with $\Lambda\leq 0$\cite{Natario:2016bay}. Here, we cannot consider the laws of thermodynamics, because one of the candidates for the final state is assumed to be a naked singularity where the black hole thermodynamics is not well defined. Then, we cannot use $dS_\text{h}=S_\text{f}-S_\text{i}$ in which $S_\text{f}$ of the naked singularity cannot be estimated. By focusing on the change in $\Delta_r$, we avoid this problem and consider broader candidates for the final state.

We have proven herein that the near-extremal K(A)dS black hole cannot be overspun, and the weak cosmic censorship conjecture is valid under the scattering of the scalar field. This is a remarkable result that is significantly different from the violation of the weak cosmic censorship conjecture by adding a particle in near-extremal black holes without considering back-reactions, such as in \cite{Toth:2011ab,Gao:2012ca,Gwak:2016gwj}. In the particle case, the effects of self-force and finite size should be considered to recover the validity of the weak cosmic censorship conjecture. Conclusively, the conjecture can be valid in both cases of particles and fields. Under the scattering of the scalar field, the near-extremal K(A)dS black hole cannot be overspun {\it without considering these effects} because we consider the infinitesimal time interval $dT$, which is not considered when adding a particle. Even if the energy and angular momentum coming into the black hole are sufficiently large to exceed the extremal condition, there are limits given in Eq.\,(\ref{eq:fluxes05}) during the time interval $dT$. Large portions of changes in the energy and the angular momentum of the black hole affect its state in several steps divided by the time interval $dT$; thus, the energy and the angular momentum cannot exceed the extremal condition. Furthermore, in Eqs.\,(\ref{eq:deltamin02}) and (\ref{eq:solomegaeff03}), although the near-extremal K(A)dS black hole is spun up by the scalar field, the black hole remains to be a near-extremal one because of the effect of the time interval. Hence, the time interval in the scalar field case plays a similar role in a series of smaller processes carrying the energy and the angular momentum in the particle case done in \cite{Chirco:2010rq}. Therefore, according to the concept of the time interval, the scattering of the scalar field makes it possible for the cosmic censorship conjecture to be valid in all the initial states of the K(A)dS black holes, such as non-extremal, near-extremal, and extremal cases.

\section{Superradiance}\label{sec6}

When the scalar field is scattered to the outer horizon of the K(A)dS black hole, the energy and the angular momentum of the scalar field are transferred to the black hole as shown by the fluxes in Eq.\,(\ref{eq:flux01}). For a given angular velocity $\Omega_\text{h}$, if the scalar field is $\omega < m\Omega_\text{h}$, the fluxes become negative, implying that the energy and the angular momentum leak from the black hole to the scattered scalar field. An interesting feature, called superradiance, exists. When we continuously maintain a scalar field of $\omega/m$ in the black hole spacetime, we can easily estimate the evolution of the black hole (i.e., the black hole will radiate the energy and angular momentum to decrease the angular velocity up to $\omega/m$ by the superradiance). Note that we set a constant state of the scalar field; hence, there is no amplification for an instability to occur. The radiation caused by the scalar field is eventually completed because the fluxes are zero in the saturated case.

We investigate herein whether the saturation $\omega/m=\Omega_\text{h}$ is completed in a finite time based on our analytical approach. By the estimation of the change in $\Omega_\text{h}$ from a step after the time interval $dT$, we can prove that the black hole will take infinite time to be saturated through the superradiance. More detailed behaviors of this phenomenon need to be studied by a numerical approach starting from a redefined Lagrangian because the mass and the angular momentum of the black hole depend on the time evolution. However, our analytical approach offers the advantage of investigation at the near-saturated point, $\omega/m\sim \Omega_h$, where such behaviors are too close to be studied by a numerical approach, which could lead to a numerical error. Note that the detailed behaviors of the overall scalar field depend on the boundary condition. The asymptotic boundary differs with the given sign of the cosmological constant; hence, the radial function of the scalar field also differs from that of the cosmological constant. Nevertheless, at the outer horizon, the radial solution of the scalar field is commonly given by Eq.\,(\ref{radialsol21}). We focus on the fluxes of the energy and the angular momentum at the outer horizon because the changes in the black hole occur in the infinitesimal time interval $dT$. The interval is too short to consider the scalar field reflected from an asymptotic boundary.

The initial angular velocity of the K(A)dS black hole is set to $\Omega_\text{h}$ of $(M_\text{B},J_\text{B})$, where the initial state of the K(A)dS black hole is not important. Furthermore, we assume that the scalar field is
\begin{align}\label{eq:kappa01}
\frac{\omega}{m}=\Omega_\text{h}-\kappa,
\end{align} 
where $\kappa\ll 1$. The initial black hole and the scalar field are nearly saturated to each other, but not fully saturated. $\omega/m$ of the scalar field is slightly smaller than the angular velocity of the black hole; thus, superradiance occurs in the initial state. The scattering changes the angular velocity to
\begin{align}
\Omega_\text{h}(M_\text{B}+dM_\text{B},J_\text{B}+dJ_\text{B})&=\Omega_\text{h}+d\Omega_\text{h},
\end{align}
where
\begin{align}
d\Omega_\text{h}&=\frac{\partial \Omega_\text{h}}{\partial M_\text{B}}dM_\text{B}+\frac{\partial \Omega_\text{h}}{\partial J_\text{B}}dJ_\text{B}+\frac{\partial \Omega_\text{h}}{\partial r_\text{h}}dr_\text{h},\quad \frac{\partial \Omega_\text{h}}{\partial M_\text{B}}=\frac{2a^3 \left(1-\frac{1}{3}r_\text{h}^2\Lambda\right)\Xi^2}{M(r_\text{h}^2+a^2)^2}-\frac{a\left(1-\frac{1}{3}r_\text{h}^2\Lambda\right)\Xi^2}{M(r_\text{h}^2+a^2)},\nonumber\\
\frac{\partial \Omega_\text{h}}{\partial J_\text{B}}&=-\frac{2a^2 \left(1-\frac{1}{3}r_\text{h}^2\Lambda\right)\Xi^2}{M(r_\text{h}^2+a^2)^2}+\frac{\left(1-\frac{1}{3}r_\text{h}^2\Lambda\right)\Xi^2}{M(r_\text{h}^2+a^2)},\quad \frac{\partial \Omega_\text{h}}{\partial r_\text{h}}=-\frac{2a r_\text{h} \Lambda}{3(r_\text{h}^2+a^2)}-\frac{2a r_\text{h} \left(1-\frac{1}{3}r_\text{h}^2\Lambda\right)}{(r_\text{h}^2+a^2)^2}.\nonumber
\end{align}
For $\omega/m=\Omega_\text{h}$, the change in the angular velocity should be $-\kappa$. The final angular velocity becomes saturated to $\Omega_\text{h}-\kappa$ of the scalar field in a finite time $dT$. From Eqs.\,(\ref{eq:fluxes05}) and (\ref{eq:kappa01}), the changes in the mass and the angular momentum of the black hole are given by
\begin{align}\label{eq:fluxes17}
dM_\text{B}=-m^2\kappa (\Omega_\text{h}-\kappa)\frac{(r_\text{h}^2+a^2)}{\Xi}dT,\quad dJ_\text{B}=-m^2 \kappa\frac{(r_\text{h}^2+a^2)}{\Xi}dT.
\end{align}
In addition, from Eq.\,(\ref{eq:radious01}), the change in the location of the outer horizon is rewritten as
\begin{align}\label{eq:drh17}
dr_\text{h}=\frac{2m^2 \kappa \left(3a(r_\text{h}^2 +a(r_\text{h}^2+a^2)\kappa)+a^3r_\text{h}^2\Lambda-\frac{9(r_\text{h}^2+a^2)^3\kappa}{-3r_\text{h}^2+3r_\text{h}^4\Lambda+a^2(3+r_\text{h}^2\Lambda)}\right)}{3(r_\text{h}^2+a^2)}dT.
\end{align}
Substituting Eqs.\,(\ref{eq:fluxes17}) and (\ref{eq:drh17}), the change in the angular velocity $d\Omega_\text{h}$ is obtained in the order of $\kappa$ as
\begin{align}\label{eq:domega17}
d\Omega_\text{h}=-\frac{2m^2  r_\text{h}^3\Xi^2dT}{(r_\text{h}^2+a^2)^2}\kappa+\mathcal{O}(\kappa^2).
\end{align}
The negative sign in the coefficient of the first order of $\kappa$ implies that the angular velocity becomes close to $\omega/m$ of the scalar field, but the magnitude of the coefficient is extremely small: $|d\Omega_\text{h}/\kappa| \ll 1$. In other words, the saturation of the angular velocity cannot be in a finite time because the angular velocity becomes slightly closer to $\omega/m$ of the scalar field for a step among the overall difference $\kappa$. After each step, the remaining value of $\kappa$ becomes smaller, but does not become zero. However, the coefficient of $\kappa$ is smaller than $1$; thus, the remaining value of $\kappa$ would be zero after infinite steps. The energy and the angular momentum fluxes become smaller as the black hole comes closer to the saturated state by the superradiance, as shown by Eq.\,(\ref{eq:fluxes17}). Therefore, the K(A)dS black hole cannot approach the saturated state with no superradiance in a finite time.

Note that our result for the superradiance can be applied to the energy of the scalar field absorbed into the K(A)dS black hole by taking $-\kappa\rightarrow +\kappa$ in Eq.\,(\ref{eq:kappa01}). The change in the angular velocity given in Eq.\,(\ref{eq:domega17}) is rewritten as
\begin{align}\label{eq:domega18}
d\Omega_\text{h}=\frac{2m^2  r_\text{h}^3\Xi^2dT}{(r_\text{h}^2+a^2)^2}\kappa+\mathcal{O}(\kappa^2).
\end{align}
That is, the K(A)dS black hole cannot saturate its angular velocity to the mode of the scalar field by energy absorption under the scattering of the scalar field.

\section{Summary}\label{sec7}

We investigated herein the validity of the weak cosmic censorship conjecture under the scattering of the scalar field in the K(A)dS black hole without imposing the laws of thermodynamics. The metric of a K(A)dS black hole is not static at the asymptotic boundary in the Boyer-Lindquist coordinates $(t,r,\theta,\phi)$. Furthermore, the laws of thermodynamics are not well defined because of the non-static boundary. For a static boundary, the Boyer-Lindquist coordinates become new coordinates $(T,r,\theta,\Phi)$ via coordinate transformation. We assume herein that the K(A)dS black hole infinitesimally changes during the infinitesimal time interval because of the transferred energy and angular momentum from their fluxes of the scalar field when the scalar field is scattered at the outer horizon of the K(A)dS black hole. The fluxes are obtained from the energy-momentum tensor of the Lagrangian with non-minimal coupling, but the non-minimal coupling and the mass of the scalar field do not contribute to the fluxes. We focus on the fluxes at the outer horizon that represent the energy and angular momentum coming into and going out of the K(A)dS black hole; therefore, the fluxes can be obtained from the solution of the scalar field at the outer horizon considering separation and normalization. The weak cosmic censorship conjecture was investigated in two cases: non-extremal and near-extremal K(A)dS black holes. For the non-extremal black holes, we showed that the Bekenstein-Hawking entropy of the K(A)dS case is irreducible under the scattering of the scalar field; hence, the horizon area proportional to the entropy cannot disappear under the scattering. In other words, the second law of thermodynamics is valid in the non-extremal K(A)dS black hole under the scattering of the scalar field. We can also induce the first law of thermodynamics from the change in the K(A)dS black hole owing to the scattering because the exact forms of the fluxes are obtained at the outer horizon. The change in the entropy is divergent at the extremal black hole, and the laws of thermodynamics are not defined in the naked singularity in the final state; thus, we investigated the weak cosmic censorship conjecture in the near-extremal black hole by the change in the minimum value of the metric component $\Delta_r$. There is a distinct result from the scalar field compared with the result obtained by adding a particle. The near-extremal K(A)dS black hole cannot be overspun because of the scattering. In other words, by changing by the scalar field, the near-extremal K(A)dS black hole is always under the extremal condition without consideration of other interactions, such as a self-force or finite size effect. This result is significantly different from the case of adding a particle. When a particle is absorbed into the black hole, the particle is treated as a portion of energy and angular momentum without the concept of time. However, the scalar field transfers its energy and angular momentum through its fluxes; hence, we need time as a constraint to limit the amount of energy and angular momentum. The changes in the mass and angular momentum of the black hole are still under the extremal condition. Thus, the weak cosmic censorship conjecture is still valid under the scattering of the scalar field. Note that this also includes the case of the extremal black hole. In addition, the K(A)dS black hole can radiate or absorb its energy and angular momentum by the scattering of the scalar field, including the superradiance; hence, we investigated whether the K(A)dS black hole can saturate its angular velocity to the mode of the scalar field $\omega=m\Omega_\text{h}$ in a finite time by the evolution caused by the scalar field. From the change in the angular velocity, we can prove that there is no way to saturate the angular velocity of the K(A)dS black hole to the mode of the scalar field; thus, the angular velocity of the K(A)dS black hole cannot be saturated to the mode of the scalar field in a finite time. Furthermore, the changes in the K(A)dS black hole are significantly different from those obtained by adding a particle because of the concept of time carried from the fluxes of the scalar field. Therefore, we expect that the invalidities of the weak cosmic censorship conjecture observed by adding a particle can be resolved under the scattering of various fields.

\vspace{10pt} 

\noindent{\bf Acknowledgments}

\noindent This work was supported by the National Research Foundation of Korea (NRF) grant funded by the Korea government (MSIT) (NRF-2018R1C1B6004349).

\end{document}